\documentclass[showpacs,preprintnumbers,amsmath,amssymb,twocolumn,superscriptaddress,prb]{revtex4}
\usepackage{graphicx}
\usepackage{amsfonts}
\usepackage{amsmath, amsthm, amssymb}
\usepackage{lmodern}
\usepackage[usenames]{color} 
\usepackage[normalem]{ulem}

\begin{document}
\title{Quantum criticality in spin chains with non-Ohmic dissipation}
\author{Iver Bakken Sperstad}
\affiliation{Department of Physics, Norwegian University of
Science and Technology, N-7491 Trondheim, Norway}
\author{Einar B. Stiansen}
\affiliation{Department of Physics, Norwegian University of
Science and Technology, N-7491 Trondheim, Norway}
\author{Asle Sudb{\o}}
\affiliation{Department of Physics, Norwegian University of
Science and Technology, N-7491 Trondheim, Norway}

\date{Received \today}
\begin{abstract}
We investigate the critical behavior of a spin chain coupled to bosonic baths characterized by a spectral density 
proportional to $\omega^s$, with $s>1$. Varying $s$ changes the effective dimension $d_\text{eff} = d + z$ of 
the system, where $z$ is the dynamical critical exponent and the number of spatial dimensions $d$ is set to one.
We consider two extreme cases of clock models, namely Ising-like and $U(1)$-symmetric ones, and find the critical 
exponents using Monte Carlo methods. The dynamical critical exponent and the anomalous scaling dimension $\eta$ 
are independent of the order parameter symmetry for all values of $s$. The dynamical critical exponent varies 
continuously from $z \approx 2$ for $s=1$ to $z=1$ for $s=2$, and the anomalous scaling dimension evolves correspondingly 
from $\eta \gtrsim 0$ to $\eta = 1/4$. The latter exponent values are readily understood from the effective dimensionality 
of the system being $d_\text{eff} \approx 3$ for $s=1$, while for $s=2$ the anomalous dimension takes the well-known exact 
value for the two-dimensional Ising and $XY$ models, since then $d_{\rm{eff}}=2$. However, a noteworthy feature is, that $z$ approaches 
unity and $\eta$ approaches $1/4$ for values of $s < 2$, while naive scaling would predict the dissipation to become irrelevant 
for $s=2$. Instead, we find that $z=1,\eta=1/4$ for $s \approx 1.75$ for both Ising-like and $U(1)$ order parameter symmetry. 
These results lead us to conjecture that for all site-dissipative $Z_q$ chains, these two exponents are related by the scaling 
relation $z = \text{max} \{(2-\eta)/s, 1 \}$. We also connect our results to quantum criticality in nondissipative spin chains 
with long-range spatial interactions. 
\end{abstract}	

\maketitle

\section{Introduction}
\label{sec:intro}

The spin-boson model\cite{Leggett-Chakravarty_dissipative_RMP,LeHur_spin-boson} (SBM) represents one of the most well-established frameworks for describing the effect of 
dissipation on a quantum system. In its simplest incarnation, it describes a two-level system coupled to an infinite number of harmonic oscillators with low-frequency spectral 
density $J(\omega) \propto \omega^s$, with Ohmic damping ($s=1$) being the most commonly studied case. Generalizations of this model include extensions to finite (spatial) 
dimensions $d>0$,\cite{Werner-Volker-Troyer-Chakravarty, Sperstad_dissipative} and models where the $Z_2$ (Ising) spin symmetry has been replaced by a higher 
symmetry.\cite{Werner-Troyer-Sachdev_dissipative_XY_chain,Vojta_Impurity_QPT} 
Extended versions of such systems may also find applications in the study of quantum critical points 
in quantum magnets and strongly correlated 
systems,\cite{Vojta_QPT_review,Vojta_Impurity_QPT,Si_Kondo_quantum_criticality,Aji-Varma_orbital_currents_PRL} 
and hence they are of considerable interest in contemporary condensed matter physics.

Another generalization is to consider non-Ohmic spectral densities ($s\neq1$), which may be relevant in the description of several different physical 
phenomena.\cite{Wagenblast-Schon_non-ohmic,Orth_quantum_Ising,Vojta_superohmic_XY,Hoyos-Vojta_disordered_dissipative,Lobos-Giamarchi_SC_wires, Lutchyn-Galitski-Refael-DasSarma_dissipation, Glossop-Ingersent_BFKM,Gamba_damping_RG}
From a  more fundamental physics point of view, the sub-Ohmic ($s < 1$) SBM and related models have been studied intensively in recent years\cite{Winter-Vojta_cont_time_dissipation,Kirchner_finite-size_dissipative,Kirchner_spin-boson} following claims that the so-called quantum-to-classical 
mapping may be violated 
in even the simplest variant of SBM for $s<1/2$.\cite{Vojta_sub-ohmic_spin-boson} Its super-Ohmic counterpart  $s > 1$ has, on the other 
hand, received relatively little attention. This may be due to the fact that the (0+1)-dimensional [(0+1)D] SBM exhibits a (quantum) phase transition only for values of 
$s \le 1$. For $d \ge 1$, however, the possibility of a phase transition arises for all $s$.

The SBM is generally described, via the quantum-to-classical mapping, by a classical ($d$+1)D spin model with long-range interactions that decay as $1/\tau^{1+s}$ 
in imaginary time $\tau$. Long-range interactions are interesting, as they allow one to increase the effective dimensionality 
continuously by tuning $s$ to lower values. In classical spin glasses, for instance, low-dimensional models with long-range interactions have been studied to infer 
properties of higher-dimensional realizations of the same systems with purely short-range interactions.\cite{Kotliar_long-range_spin_glass, Katzgraber_long-range_spin_glass, Sharma-Young_long-range_Heisenberg} In quantum models, the effective dimensionality is expressed by $d_\text{eff}=d+z$, with $z$ being the dynamical critical exponent 
defined from the divergence of the correlation time 
$\xi_\tau \sim \xi^z$, where $\xi$ is the spatial correlation length. At a second-order phase transition, we have in standard notation $\xi \sim |K-K_c|^{-\nu}$ as the coupling 
parameter $K$ approaches its critical value $K_c$. The presence of dissipation in general causes $z$ to deviate from the value $z=1$, with a naive scaling estimate $z_0=2/s$.\cite{Hertz_quantum_critical} Although this result is exact in mean-field theory ($d_\text{eff} \geq 4$), deviations may appear when decreasing 
$d_\text{eff}$. For the Ohmic case, it is known\cite{Pankov-Sachdev_z2} that $z$ obeys the scaling law $z = z_0 - \eta$, where $\eta$ 
in general denotes the anomalous scaling dimension at the transition point to a disordered state. Below $d_\text{eff}=4$ one has $\eta > 0$, and previous
work\cite{Werner-Volker-Troyer-Chakravarty} on $s=1$ for $d=1$ found $\eta \approx 0.015$ and $z \approx 1.985$.

One issue we address in this paper is how the exponents $z$ and $\eta$ evolve as one varies the dissipation parameter $s >1$. For the Ohmic case considered previously, 
the deviations from naive scaling (i.e., from $z = z_0$) are barely significant due to the small value of $\eta$ when $d_\text{eff} \approx 3$. This deviation should become 
more noticeable as the effective dimensionality decreases, although one cannot expect the relation $z = z_0 - \eta$, valid for $s=1$, to hold also for larger $s$. In the 
limit $d+z \to 2$, the anomalous dimension might be expected to approach the relatively large value $\eta = 1/4$, which it takes for both the two-dimensional (2D) Ising and 2D $XY$ model. 
A related issue is the value of $s$ beyond which the dissipation term is irrelevant in the renormalization group sense, giving $z=1$. Naive scaling indicates that $z = 1$ 
for $s \geq 2$, but as $z$ is likely to decrease faster than $z_0 = 2/s$ as $s$ increases, dissipation might turn irrelevant for a value of $s$ smaller than 2.

Another issue which we address is how the critical exponents, in particular $z$ and $\eta$, depend on the symmetry of the order parameter. In the limit 
$s=1$, there is no significant difference between the values of $\eta$ (and thereby $z$) for discrete and continuous order parameter fields, 
\cite{Werner-Volker-Troyer-Chakravarty,Werner-Troyer-Sachdev_dissipative_XY_chain} but it is not inconceivable that such a difference becomes noticeable 
for lower effective dimensions, that is, as $s$ is increased.

In order to answer these questions and to study a class of dissipative models for which relatively little is known precisely, we present results from Monte Carlo simulations 
on both $XY$ and Ising-like spin chains with non-Ohmic dissipation. In both cases, we consider super-Ohmic dissipation, which for the $XY$ case allows us to interpolate between the 
universality class describing the three-dimensional (3D) $XY$ model and the very different Berezinskii-Kosterlitz-Thouless (BKT) criticality of the 2D $XY$ model. The models are presented in the 
next section, where we also describe the finite-size scaling procedure used to find the critical exponents. The dependence of these exponents on $s$ are presented and discussed 
in Sec. \ref{sec:results}, before we give a summary of our findings in Sec. \ref{sec:summary}.

\section{Model and finite-size scaling methods}
\label{sec:method}

The starting point of the models we consider may be taken as a general (1+1)D $\phi^4$-type quantum field theory of an $O(N)$ order parameter field $\phi$. Including dissipation, 
the Fourier transform of its inverse bare propagator is of the form $q^2 + \omega^2 + |\omega|^s$, where the damping term $\propto |\omega|^s$ arises from the coupling of the 
field to baths of harmonic oscillators\cite{Caldeira-Leggett} with a low-frequency power-law 
spectral density characterized by the exponent $s$. 

Parameterizing the order parameter field of such a $N=2$ quantum rotor model by an angle variable $\theta$, we may formulate the discretized action as
\begin{align}
  \label{eq:action}
  S &= -K \sum_{x=1}^L \sum_{\tau=1}^{L_\tau}\cos( \theta_{x,\tau} - \theta_{x+1,\tau})  \\ \nonumber
  &-K_\tau \sum_{x=1}^L \sum_{\tau=1}^{L_\tau}\cos( \theta_{x,\tau} - \theta_{x,\tau+1})  \\ \nonumber
  &- \frac{\alpha}{2}\sum_{x=1}^L \sum_{\tau\neq\tau'}^{L_\tau}\left(\frac{\pi}{L_\tau}\right)^{1+s}
\frac{\cos{(\theta_{x,\tau} - \theta_{x,\tau'})}}{\sin^{1+s}(\frac{\pi}{L_\tau}|\tau-\tau'|)}
\end{align}
on a quadratic $L \times L_\tau$ lattice. Above, $K$ is the spatial coupling constant to be varied, whereas the quantum coupling constant $K_\tau$ and the dissipation strength $\alpha$ are taken as fixed values during the simulations.

In order to study both continuous and discrete symmetry of the order parameter field, we consider two possible domains of the angle variables: $U(1)$ symmetry is equivalent with $\theta \in \lbrack 0,2\pi \rangle$, and for a discrete symmetry ($Z_4$), we choose to enforce the restriction $\theta \in \lbrace 0, \frac{\pi}{2}, \pi, \frac{3\pi}{2} \rbrace$. We refer to the former as the $XY$ model and to the latter as the  $Z_4$ model. Such a $Z_4$ model will be in the same universality class as a corresponding $Z_2$ (Ising) model, which is why we refer to this model as Ising-like. 
This equivalence is easily shown
using the substitution $\cos{\theta_{x,\tau}} = (\sigma_{x,\tau} + \mu_{x,\tau})/2$ and $\sin{\theta_{x,\tau}} = (\sigma_{x,\tau} - \mu_{x,\tau})/2$ to 
rewrite the action as that of two Ising models in terms of decoupled Ising spins $\sigma$ and $\mu$. 
The order parameter for both symmetry variants is defined as $m = (L L_\tau)^{-1} \sum_{x,\tau}{\exp{(\mathrm{i} \theta_{x,\tau})}}$ in the standard manner.

When determining critical exponents of a quantum system using finite-size scaling (FSS), the system dimensions $L$, $L_\tau$ used have to be chosen such that they respect 
the system anisotropy reflected by the dynamical critical exponent, $L_\tau \propto L^z$. This is a problem when we do not know dynamical critical exponents \textit{a priori}, 
and one usually has to first determine $z$ by simulating several values of $L_\tau$ for each $L$, before running new simulations with $L / L_\tau^z$ fixed. We circumvent this 
problem by using the same data to determine $z$ and to evaluate the FSS observables, by interpolating data for multiple $L_\tau$ values to $L_\tau = L_\tau^\ast(L)$.
Here, $L_\tau^\ast$ is a characteristic temporal system size found for each spatial system size $L$, as explained below, and it is assumed that $L_\tau^\ast \propto L^z$. This 
has the advantage that one (along with $z$) can find all other critical exponents simultaneously, utilizing all (or most of) the data generated. Furthermore, we are also able 
to appropriately take the uncertainty in $z$ into account when finding the uncertainty in the other exponents, by repeating the entire procedure for a number of jackknife 
bins based on the original data.

The procedure to find $z$ is explained in more detail in, for example, Ref. ~\onlinecite{Sperstad_dissipative}, and is based on the Binder ratio
\begin{equation}
	Q = \frac{ \langle m^4 \rangle}{\langle m^2 \rangle^2} = \mathcal{Q}(|K-K_c|L^{1/\nu},L_\tau/L^z),
\end{equation}
where brackets $\langle \ldots \rangle$ indicate ensemble averages and $\mathcal{Q}$ is a universal scaling function. The characteristic values $L_\tau^\ast(L)$ are found 
from the minima of $Q$ as a function of $L_\tau$ for a given $L$, and the critical coupling $K_c$ is found from the crossing points of these minima as a function of $K$. 
The correlation length exponent $\nu$ is determined through finite-size scaling of the related quantity
\begin{equation}
	\frac{\left(\partial \langle m^2 \rangle / \partial K \right)^2 }{\partial \langle m^4 \rangle / \partial K} \propto L^{1/\nu},
\end{equation}
where the derivatives are calculated by $\partial \langle m^n \rangle / \partial K = \langle E_x \rangle \langle m^n \rangle - \langle E_x m^n \rangle$, with $E_x = -\sum_{x,\tau} \cos( \theta_{x,\tau} - \theta_{x+1,\tau})$. 

To extract critical exponents $\beta$ and $\gamma$, we use the usual FSS forms for the magnetization
\begin{equation}
	\label{eq:magn}
	\langle |m| \rangle \propto L^{-\beta/\nu}
\end{equation}
and the magnetic susceptibility
\begin{equation}
	\label{eq:susc}
	\chi = L L_\tau \langle m^2 \rangle \propto L^{\gamma/\nu},
\end{equation}	
respectively.
The anomalous dimension $\eta$ is then found from the scaling relation $\eta =2 - \gamma/\nu$. We have also checked that the value of $\eta$ obtained from the susceptibility 
data is in correspondence with that obtained (through $z+\eta$) from the critical two-point correlation function of the order parameter field, $G(L/2) \propto L^{2-d-z-\eta}$.
All of the above observables are evaluated at $L_\tau = L_\tau^\ast$, and we are careful to only use system sizes $L_\tau$ relatively close to $L_\tau^\ast$ 
in the interpolation. Using a polynomial fit of as low order as 3 works very well in most cases, although more care must be taken when extracting $\nu$.

The error estimates we report are jackknife estimates of statistical errors only, 
but include contributions from the uncertainty in $L_\tau^\ast$ and the critical coupling $K_c$. 
The value of $K_c$ is in general extrapolated from the scaling form 
$K_c^\ast(L) = K_c + c L^{-\omega'}$ for the crossing points $K_c^\ast(L)$ of 
$Q(K,L_\tau=L_\tau^\ast)$ for adjacent system sizes $L$.
For regions of $s$ where the crossing points coincide and the extrapolation procedure breaks down, we base the estimate of $K_c$ on the largest values of $L$.
When extracting critical exponents, we make sure to use system sizes large enough for the above mentioned FSS forms to be valid. 
Possible corrections to scaling are discussed below.
We note in particular that we initially assume a second-order phase transition for all values 
of $s$ we use, so critical exponents obtained in the case of a BKT transition should be regarded as effective exponents only.
The special case of $s \approx 2$ for the $XY$ model is therefore re-examined separately in Sec. \ref{sec:BKT}.
Corrections to the scaling form $L_\tau^\ast \propto L^z$ are discussed for a special case in Sec. \ref{sec:LR-SR}.

The Monte Carlo simulations are performed using a Wolff cluster 
algorithm\cite{Wolff} for long-range interactions.\cite{Luijten-Blote} 
The results are obtained using an implementation of the {\scshape Mersenne Twister}\cite{Mersenne_Twister} 
random number generator, but other random number generators produce consistent results. 
Ferrenberg-Swendsen reweighting techniques\cite{Ferrenberg-Swendsen} were applied to the data.
For the simulations of the $XY$ case, we use a model with $Z_{32}$ 
symmetry to emulate the continuous $U(1)$ symmetry.

\section{Results and discussion}
\label{sec:results}

When extracting critical exponents for the model we consider in this paper, 
we anticipate that the only parameter in Eq. \eqref{eq:action} relevant in determining the universality class is the interaction decay exponent $s$.
(The values we present Monte Carlo results for are $s =1$, 1.25, 1.5, 1.625, 1.75, 1.875, and 2.)
Nevertheless, we also find that the corrections to scaling are strongly affected by the 
strength of the dissipation term as quantified by $\alpha$, for finite systems. 
In order to minimize finite-size effects and ensure a relatively fast onset of asymptotic 
values of the exponents, a specific value of $\alpha$ could in principle be tailored to each 
value of $s$.\cite{Hasenbusch_improved_actions}
Instead of adjusting $\alpha$ for each individual value of the 
decay exponent $s$, we have divided the span of $s$ values into two regions where we have applied different sets of coupling constants. For $s > 1.625$ where we expect the 
dissipation term to be weakly relevant (in the sense of a small correction-to-scaling 
exponent) or even irrelevant, we set the coupling values according to $\alpha=0.1$ and 
$K_\tau = -\ln{(\tanh{\frac{1}{2}})} \approx 0.7719$. 
For $s \leq 1.5$ we find that it is more appropriate 
to choose a larger value of $\alpha$ while reducing $K_\tau$ 
in order to observe a rapid finite-size crossover to the asymptotic exponents. In this region we use $\alpha=0.5$ and $K_\tau=0.4$. For the intermediate value $s=1.625$, we 
use $\alpha=0.3$ and $K_\tau=0.4$. We can easily confirm for the smallest values of $s$ that the universality class does not depend on the value of $\alpha$, but corrections 
to scaling makes this harder for larger $s$, as discussed in Sec. \ref{sec:LR-SR}.

\subsection{Results for the critical exponents}

\begin{figure}[tbh]
	\includegraphics[width=0.4\textwidth]{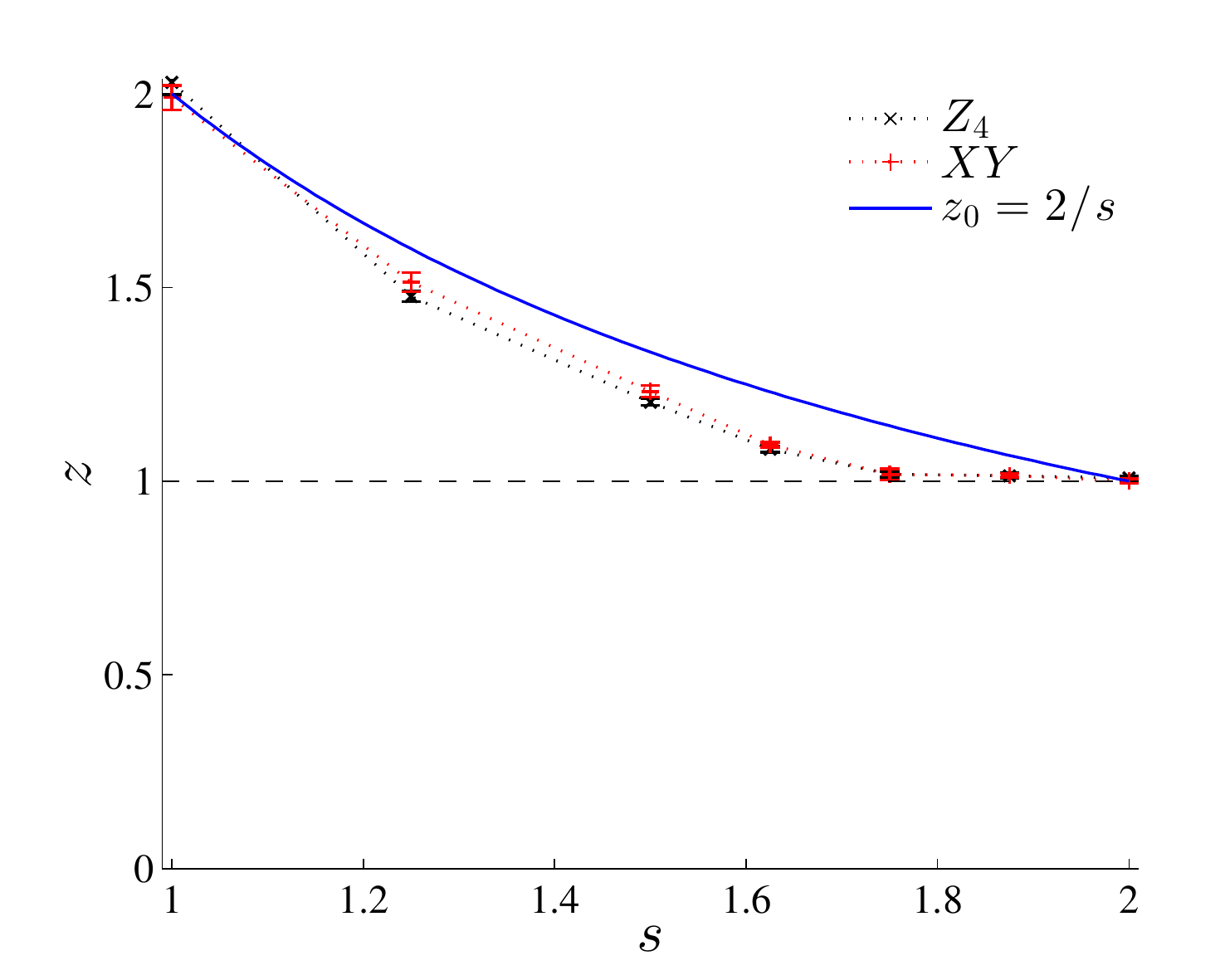}
	\caption{Dynamical critical exponent $z$ as a function of $s$ for the $Z_4$ and the $XY$ model. 
The naive scaling estimate $z_0$ (the solid curve) does not coincide with the calculated $z$ for values of $s$ other than the integer-valued end points of our 
span of $s$ values.}
	\label{fig:z_vs_s}
\end{figure}

In Fig. \ref{fig:z_vs_s} we present the dynamical critical exponent $z$ as a function of $s$. A notable feature of the results is the similarity between the two order 
parameter symmetries. To the accuracy of our simulations, there is essentially no difference between the continuous $U(1)$ symmetry and the discrete $Z_4$ symmetry. Also, 
the calculated $z$ values 
do not conform to the scaling estimate $z_0=2/s$, but 
instead fall off faster for increasing $s$ than expected from naive scaling.

\begin{figure}[tbh]
	\includegraphics[width=0.4\textwidth]{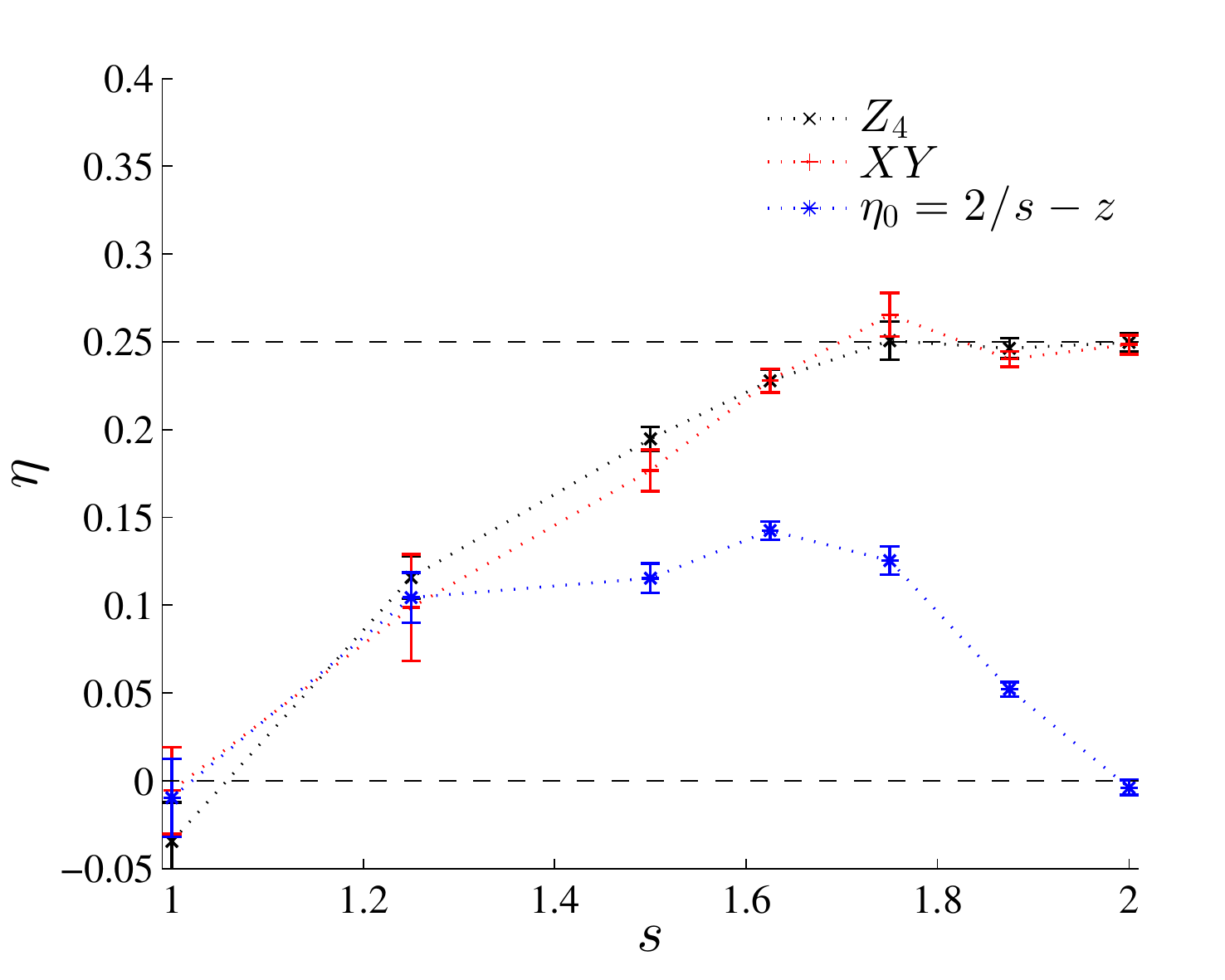}
	\caption{Anomalous scaling dimension $\eta$ as a function of $s$ for the $Z_4$ and the $XY$ model. $\eta_0$ indicates the discrepancy between the naive scaling estimate 
$z_0=2/s$ and the actually calculated value of the dynamical critical exponent $z$ (based on the mean for the $Z_4$ and $XY$ model).}
	\label{fig:eta_vs_s}
\end{figure}

We present the evolution of $\eta$ as a function of $s$ in Fig. \ref{fig:eta_vs_s}. Again, we find coinciding values for the two order parameter symmetries. For 
both $Z_4$ and $U(1)$, $\eta$ increases steadily with decreasing effective dimension of the system. 
Also shown in Fig. \ref{fig:eta_vs_s} is the quantity $\eta_0 = 2/s - z$, 
which quantifies the difference between the naive scaling estimate $z_0$ and the calculated $z$. For $s\leq 1.25$ the evolution of $\eta_0$ 
closely follows the calculated values of $\eta$, making the scaling relation $z=2/s - \eta$ a fair approximation also for $s \gtrsim 1$. For larger values of $s$, 
however, this scaling relation has clearly broken down, as the values of $z$ again approach the naive estimate as $s \to 2$. In this limit, $\eta$ approaches the 
value $\eta=1/4$, which is expected for both the 2D Ising model at the critical point, as well as for the 2D $XY$ model at the critical end point.

\begin{figure}[bth]
	\includegraphics[width=0.4\textwidth]{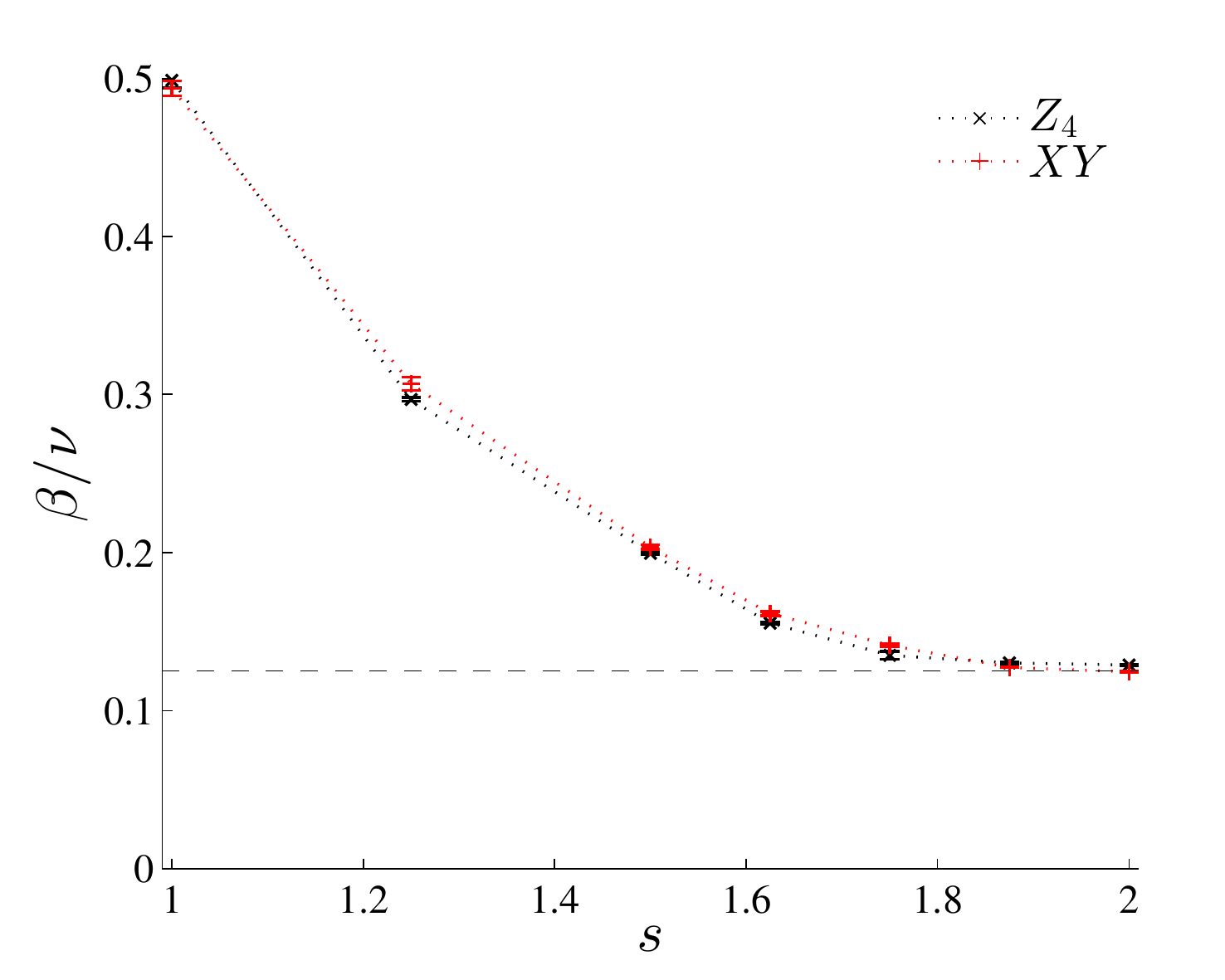}
	\caption{Critical exponent ratio $\beta/\nu$ as a function of $s$. This ratio appears to be independent of order parameter symmetry and is also well defined 
    in the limit of large $s$. The dashed line represents $\beta/\nu = 1/8$.}
	\label{fig:betaNu_vs_s}
\end{figure} 

Next, we turn to the remaining critical exponents. Figure \ref{fig:betaNu_vs_s} shows the results for the ratio $\beta/\nu$ as obtained from the magnetization.
We do not show the ratio $\gamma/\nu$, although its behavior is easily inferred from Fig. \ref{fig:eta_vs_s} and the relation $\gamma/\nu = 2 - \eta$.
Again, the FSS exponent seems to take essentially the same values for the $XY$ model as for the $Z_4$ model. This is also the case for $s\rightarrow 2$, where we expect 
the dissipation to be irrelevant so that the effective dimensionality is reduced to $d_{\mathrm{eff}} = 2$.
For the $XY$ model, the $U(1)$ symmetry of the variables then cannot be spontaneously broken, and the strong-coupling phase of the model features only quasi-long-range 
order (QLRO). Nonetheless, the system develops a finite magnetization $m$ as a finite-size effect, with a well-defined FSS exponent. The value 
$\beta/\nu \approx 0.125 = 1/8$ of this exponent when $s=2$ (as well as the corresponding susceptibility ratio $\gamma/\nu \approx 7/4$) is also found for the classical 2D 
$XY$ model and is, incidentally, the same as the corresponding ratio in the 2D Ising model. We discuss this issue in more detail in Sec. \ref{sec:symmetry}.

\begin{figure}[tbh]
	\includegraphics[width=0.4\textwidth]{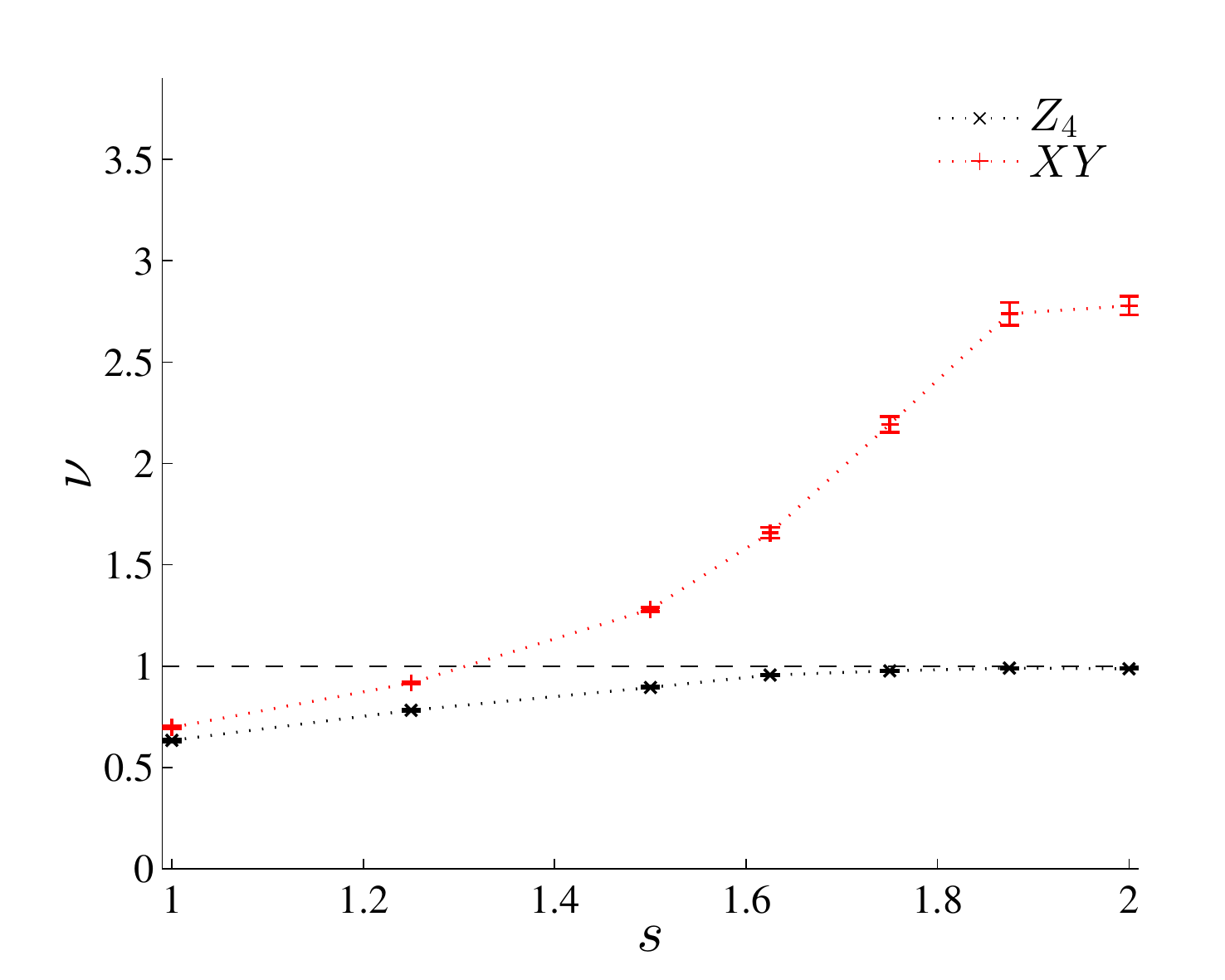}
	\caption{Correlation length exponent $\nu$ as a function of $s$. As the dissipation term becomes more short ranged with larger $s$, the exponent for the $Z_4$ 
     model approaches the 2D Ising value $\nu=1$. In the $XY$ case, $\nu$ is expected to diverge in the limit $d_{\mathrm{eff}} \to 2$. The results (obtained for finite $L$) presented for the largest
     values of $s$ should therefore be regarded only as effective exponents.}
	\label{fig:nu_vs_s}
\end{figure}

\begin{figure}[bth]
	\includegraphics[width=0.4\textwidth]{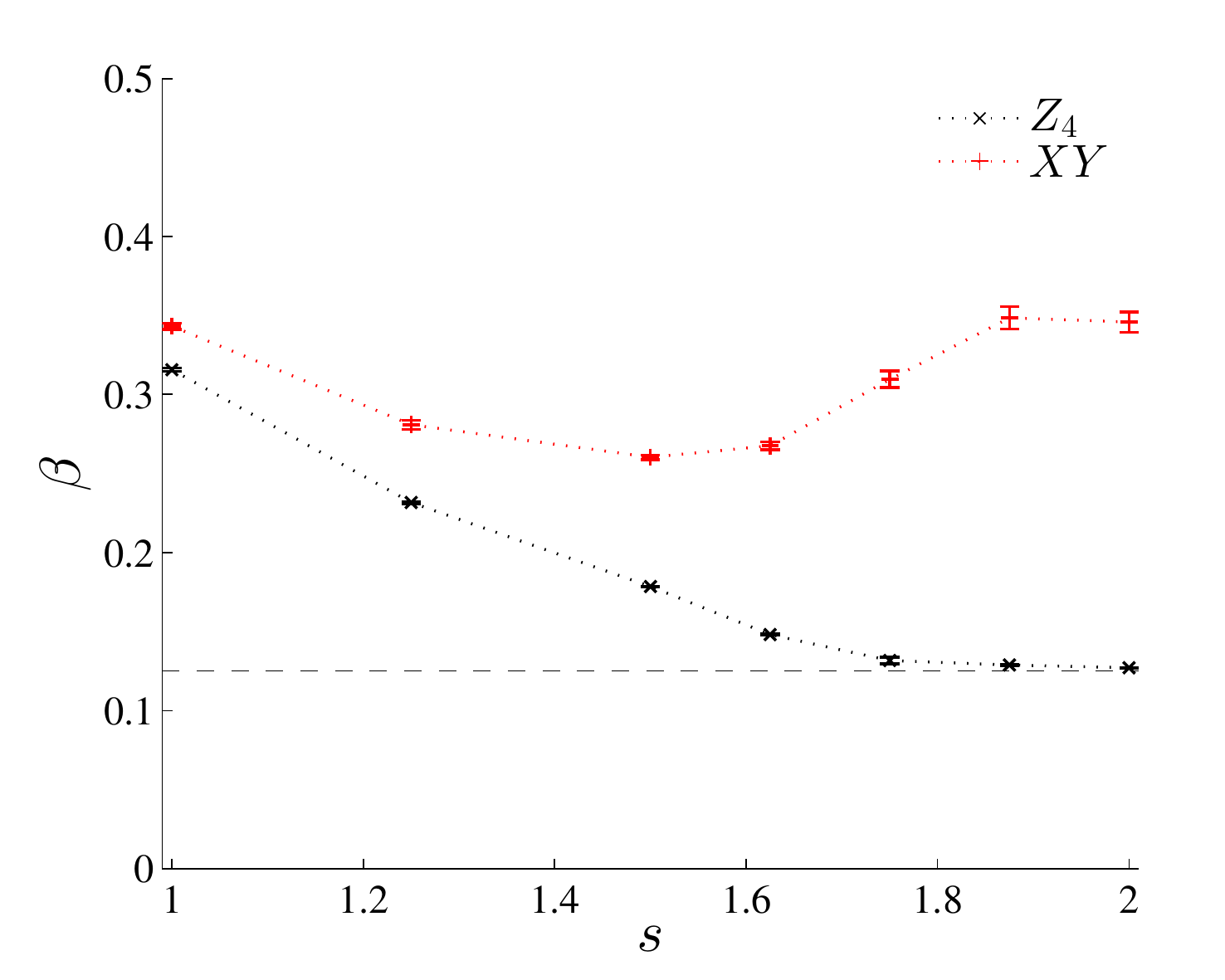}
	\caption{Critical exponent $\beta$ as a function of $s$. For the $Z_4$ model, $\beta$ evolves smoothly from the 3D Ising to the 2D Ising limit 
     as $s$ is increased from 1 to 2. The $XY$ result, on the other hand, starts out near the 3D $XY$ value for $s=1$ and features a nonmonotic evolution 
     of $\beta $ with $s$, with a divergent $\beta$ in the limit of large $s$; cf. Fig. \ref{fig:nu_vs_s}.}
	\label{fig:beta_vs_s}
\end{figure}

The correlation length exponent $\nu$ is shown in Fig. \ref{fig:nu_vs_s}, while the critical exponent $\beta$ is shown in Fig. \ref{fig:beta_vs_s}. 
We do not show the results for the exponent $\gamma$ here, but its behavior is qualitatively very similar to that of the exponent $\nu$. In the $Z_4$ 
case, both exponents start out close to the 3D Ising limit for $s=1$, and approach the 2D Ising limit indicated by the dashed line when $s \to 2$. 
Consider now the $XY$ case. When $s=1$, these exponents take on values close to those of the 3D $XY$ model. However, the exponents $\beta$, $\nu$ and $\gamma$ are 
not well defined when $d_{\mathrm{eff}} = d+z=2$, as their values are formally infinite at a transition separating a disordered phase and a QLRO phase. 
This is the case when $s=2$. Our FSS analysis for these exponents, which presupposes a second order phase transition, is strictly speaking not applicable 
to the BKT transition.  The resulting (effective) exponents $\beta$, $\nu$ and $\gamma$ appear to diverge as $L \rightarrow \infty$ close to $s=2$. Note 
that although $\xi$ is exponentially divergent at a BKT transition, we may still define $z$ through the relation $\xi_\tau \propto \xi^z$.

Another observation in the $U(1)$ case is that while the combination $\beta/\nu$ is monotonically decreasing with increasing $s$, $\beta$ itself is 
exhibiting a non-monotonic evolution as a function of $s$. The value of  $\beta$ is at first decreasing as the increasing value of $s$ drives the 
system away from the 3D behavior, just as for the $Z_4$ case. However, as mentioned above, $\beta$ is divergent in the 2D $XY$ limit, and the reduction 
of $\beta$ is therefore reversed at an intermediate value of $s$

\subsection{Berezinskii-Kosterlitz-Thouless phase transition and the helicity modulus}
\label{sec:BKT}

The helicity modulus
\begin{align}
\Upsilon &= \frac{1}{L L_\tau}\left\langle\sum_{x=1}^L \sum_{\tau=1}^{L_\tau}\cos(\theta_{x,\tau} - \theta_{x+1,\tau}) \right\rangle \\ \nonumber
& - \frac{K}{L L_\tau} \left\langle \left(\sum_{x=1}^L \sum_{\tau=1}^{L_\tau} \sin(\theta_{x,\tau} - \theta_{x+1,\tau}) \right )^2 \right\rangle,
\end{align}
is expected to scale as $\Upsilon \propto L^\kappa$ at a critical point where a $U(1)$ symmetry is spontaneously broken, with $\kappa \equiv 2\beta/\nu - \eta$.
For a 2D $XY$ model, however, the exponent $\kappa$ is exactly zero, reflecting the fact that at the BKT phase transition, the helicity modulus jumps to a finite value 
with logarithmic finite-size corrections. By direct comparison of the calculated values of $\Upsilon$ for $s=2$ and the scaling form expected for a BKT 
transition,\cite{Weber-Minnhagen_2DXY} unambiguous conclusions regarding the universality class of the phase transition at $s=2$ could not be drawn. The 
presence of the presumably irrelevant dissipation term is still effective in driving the system away from BKT-type criticality at all but the very largest 
system sizes. In practice, the logarithmic scaling analysis\cite{Weber-Minnhagen_2DXY} is usually best suited for small to moderate system sizes. Consequently, 
instead of scaling the helicity modulus directly, we resort to calculating $\kappa$ via other observables, and find that $2\beta/\nu - \eta = 0$ within statistical 
uncertainty for $s = 2$. Moreover, $2\beta/\nu - \eta$ is very close to zero for all $s \geq 1.75$. For even smaller values of $s$ (where direct scaling of $\Upsilon$ 
is more reliable), we have confirmed that the scaling law $\kappa = 2\beta/\nu - \eta$ is valid also in the presence of dissipation. This scaling form is also 
equivalent to $\kappa = d_\text{eff}-2$. Thus, the helicity modulus vanishes continuously as $K \to K_c^{+}$, provided $d_\text{eff} = d + z > 2$. The above 
equivalence assumes that hyperscaling is valid, and we have confirmed this validity for all values of $s$.

\subsection{Boundary between long-range and short-range critical behavior}
\label{sec:LR-SR}

From Figs. \ref{fig:z_vs_s} to \ref{fig:beta_vs_s}, it is evident that all critical exponents are very close to their short-range values for $s \gtrsim 1.75$. The naive 
scaling estimate places the boundary at which the dissipation term becomes irrelevant at $s=2$. For classical models with (isotropic) long-range interactions decaying with 
distance $r$ as $1/r^{1+s}$, it has long been debated\cite{Fisher_long-range_exponents,Sak_long-range} whether the models feature the exponents of the corresponding 
short-range model already when $s$ exceeds a  value $s^\ast = 2 - \eta_\text{SR}$. Here, $\eta_\text{SR}$ denotes the anomalous dimension of the short-range model. Using 
large-scale Monte Carlo simulations, it has been shown\cite{Luijten-Blote_LR-SR} that for the long-range 2D Ising model, the anomalous dimension follows the conjectured 
exact\cite{Fisher_long-range_exponents} relation $\eta = 2 - s$ for $s < 1.75$, but that $\eta = 0.25 = \eta_\text{SR}$ for $s > s^\ast = 1.75$. Although the long-range 
interaction of the dissipative quantum models we consider is highly anisotropic, in contrast to the isotropic classical long-ranged models, it is plausible that also 
in these models the threshold value of $s$ beyond which dissipation is irrelevant is reduced from $s=2$ to some lower value. 

In order to  establish this boundary more accurately also for the present case of anisotropic interactions, we have performed a more careful analysis of the case $s=1.875$ for the $Z_4$ model. Including corrections to scaling, using the ansatz $L_\tau^\ast = a L^z (1 + b L^{-\omega})$, we find  $z=1.002(11)$.
Hence, the decay exponent $s=1.875$ may serve as an upper bound for the boundary value $s^\ast$ 
necessary to render the dissipation term effectively short-ranged. This, in turn, would render the system effectively Lorentz invariant with $z=1$. To get 
the statistics required to include corrections to scaling in a stable manner, we included three different values of the dissipation 
strength $\alpha$ in a joint fit. This also provides an a posteriori justification of the choice of lower values of $\alpha$ for higher values of $s$.\cite{footnote} 
Probably due to logarithmic corrections expected at the presumed boundary value $s^\ast = 1.75$,\cite{Luijten-Blote_LR-SR}
we are not able to acquire the same level of accuracy for this value of $s$. 
Therefore we cannot rule out that the dissipation term is rendered
effectively short ranged at some other value $s^\ast \in \langle 1.75, 1.875 \rangle$. An exceedingly slow crossover to asymptotic critical exponents for values $s \approx s^\ast$ can 
conceivably be understood from the competition between the fixed point corresponding to short-range (Lorentz-invariant) criticality and the fixed point corresponding 
to long-range (dissipation-dominated) criticality.

We close this section with a remark on the evolution of the anomalous dimension.
In the quantum dissipative model we have studied, the anomalous dimension increases for increasing $s$. This is a consequence of the effective dimensionality
$d_{\mathrm{eff}}=d+z$ decreasing with increasing $s$. Lowering the dimensionality from the upper critical dimension, where $\eta=0$, tends to increase $\eta$. This is quite 
different from the situation encountered in classical models with isotropic long-range interactions. Classical models with short-range interactions and an action of the form 
$S \sim q^2 \phi_q \phi_{-q}$ (where $\phi_q$ is an appropriate order-parameter field) have propagators $G(q) \sim 1/|q|^{2-\eta}$. The corresponding long-range models with 
an action of the form $S \sim |q|^s \phi_q \phi_{-q}$ have propagators $G(q) \sim 1/|q|^s$, when $s < 2 - \eta_\text{SR}$. One may now, as is customarily done in the literature 
on long-range classical isotropic models, define an effective anomalous scaling dimension for such systems by comparing with the corresponding expression for the short-range 
case, finding $\eta=2-s$, which decreases with increasing $s$. This relation is best viewed as a result of somewhat artificially imposing the standard scaling form of a 
propagator for short-range systems ($1/|q|^{2-\eta}$) on the form of the propagator for systems with long-range interactions, $1/|q|^s$.

\subsection{Scaling relation between $z$ and $\eta$}
\label{sec:scaling}

\begin{figure}[bth]
	\includegraphics[width=0.4\textwidth]{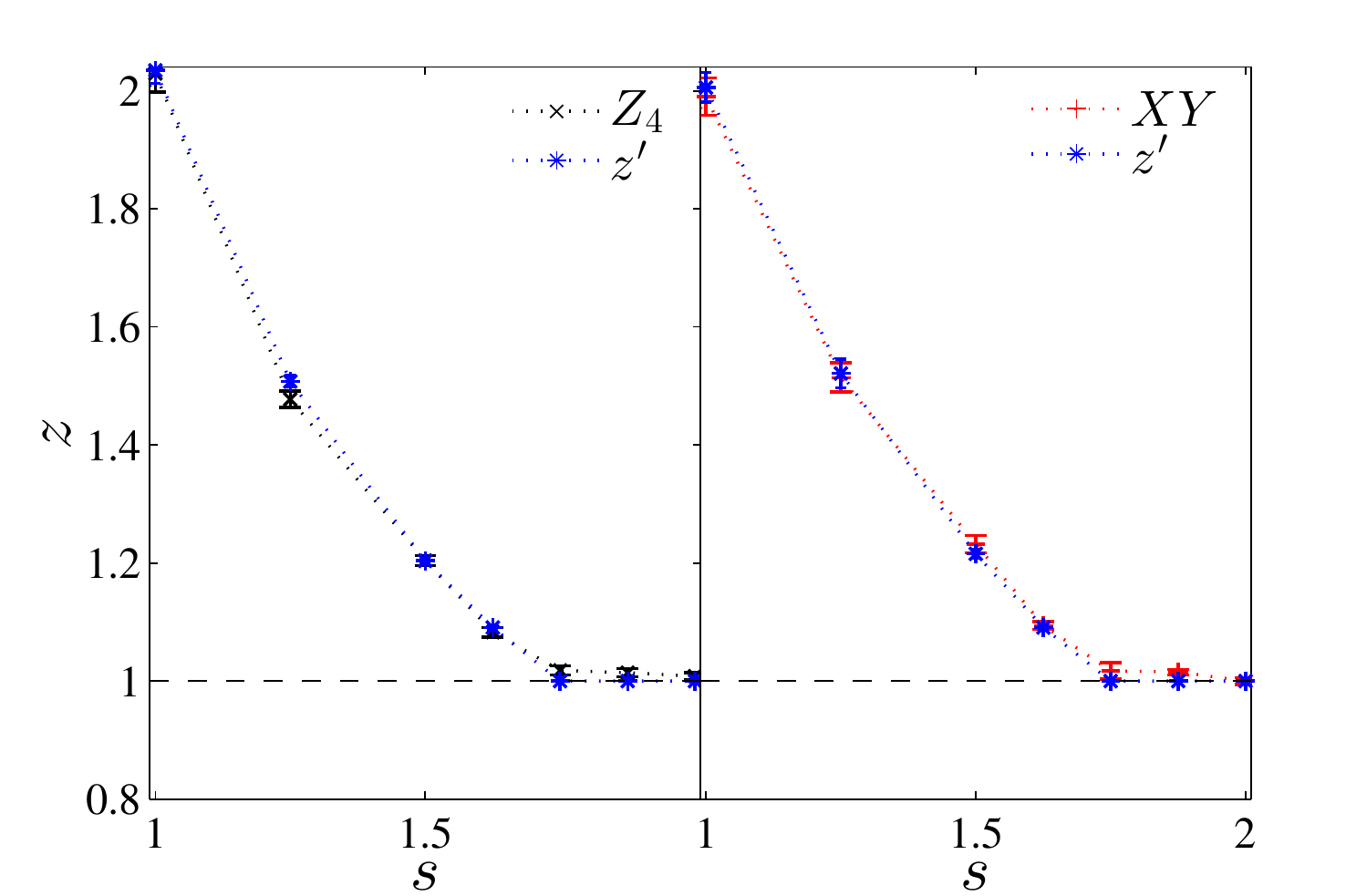}
	\caption{The dynamical critical exponent $z$ as in Fig. \ref{fig:z_vs_s}, but compared with the scaling estimate $z' = (2-\eta)/s$. The values of $\eta$ 
      used in the scaling estimates are the same as reported in Fig. \ref{fig:eta_vs_s}, with the left panel corresponding to the $Z_4$ model and the right 
      panel to the $XY$ model.}
	\label{fig:zScale_vs_s}
\end{figure}

In Fig. \ref{fig:eta_vs_s}, we demonstrated how the scaling relation $z = z_0 - \eta$ cannot be valid except close to the Ohmic limit $s=1$ and that 
$\eta \approx 1/4$ for all $s \geq 1.75$. Moreover, from our numerics we think it is likely that $z(s)=1$ for $s \geq 1.75$. A scaling relation between $z$ 
and $\eta$ which would fit well with these observations is 
\begin{equation}
	z = \text{max} \left\{ \frac{2-\eta}{s}, 1 \right\}. 
\label{eq:z_eta_Ansatz}
\end{equation}
The scaling relation $z = (2-\eta)/s$ has  been suggested previously in Ref. ~\onlinecite{Gamba_damping_RG} in the context of a damped nonlinear $\sigma$ model. In 
Fig. \ref{fig:zScale_vs_s}, we show the same data for the dynamical exponent $z$ as in Fig. \ref{fig:z_vs_s} but compared with the ansatz \eqref{eq:z_eta_Ansatz} 
instead of the naive scaling estimate $z_0 = 2/s$. Although there are probably still some corrections to finite-size scaling, 
Eq. \eqref{eq:z_eta_Ansatz} seems to fit the data far better than the alternatives. 

We next provide a heuristic argument for why the scaling relation \eqref{eq:z_eta_Ansatz} may be reasonable for $s < 2 - \eta_\text{SR}$. Building on the arguments 
in Sec. \ref{sec:LR-SR} for classical isotropic long-range-interacting systems, we take as a starting point that a dissipative quantum model with action of the form 
$S \sim (q^2 + |\omega|^s) \phi_{q,\omega} \phi_{-q,-\omega}$ can be viewed as an anisotropic long-range-interacting system. Introducing the suitably chosen 
frequency coordinate $\tilde{\omega}^s = q^2 + |\omega|^s$, the propagator takes the isotropic form 
$G(\tilde{\omega}) \sim 1/|\tilde{\omega}|^s$.  Recall that in the quantum case, the anomalous scaling dimension $\eta$  is defined from the 
\emph{spatial} correlation function $G(x) \sim 1/x^{d+z-2+\eta} \equiv 1/x^{\theta_x}$. To find $\eta$, we Fourier transform the propagator to obtain the 
imaginary-time correlation function. In terms of the frequency coordinate $\tilde{\omega}$, the system effectively has $d'_\text{eff} = 1 + d/z$ dimensions. Therefore, 
the correlation function decay exponent in terms of the isotropic space-time coordinate $\tilde{\tau} \equiv (\tau^2 + x^{2z})^{1/2}$ would be 
$\theta_{\tilde{\tau}} =	(z + d)/z - s$. Comparing with the imaginary-time decay exponent $\theta_\tau = \theta_x/z = (d+z-2+\eta)/z$, we find $sz = 2 - \eta$, 
which is equivalent to Eq. \eqref{eq:z_eta_Ansatz}.

Finally, we point out that our results for the scaling relation \eqref{eq:z_eta_Ansatz} for dissipative models should also have relevance for \emph{nondissipative} 
quantum spin chains with long-range \emph{spatial} interactions.\cite{Dutta_spatial_long-range, Laflorencie-Affleck-Berciu_long-range_Heisenberg} One arrives at 
exactly such a model by simply interchanging the $x$ and $\tau$ coordinates of the action we have considered. The dynamical critical exponent of this model is given by $z' = 1/z = s/(2-\eta)$, with the quantity $\eta(s)$ evolving as shown in Fig. \ref{fig:eta_vs_s}. This quantity will, however, not be identical to 
the anomalous scaling dimension of the model, $\eta'$, which is given by the classical result $\eta' = 2 - s$. Hence, there exists no independent scaling 
relation between the dynamical critical exponent $z'$ and the anomalous dimension $\eta'$ for a nondissipative quantum spin chain with long-range interactions. 

\subsection{Dependence on symmetry}
\label{sec:symmetry}

For $s=1$, we have $z\approx 2$, $d_{\mathrm{eff}} \approx 3$, while for $s=2$, we have $z = 1$, $d_{\mathrm{eff}} = 2$. For these two cases, it is known
either analytically or numerically that the exponent $\eta$ is very similar for the Ising and $XY$ models. \cite{Pelissetto_critical_review} There appears to be no particular 
deep reason for this. For instance, the well-known value $\eta=1/4$ comes about for completely different reasons in the 2D Ising and  2D $XY$ models, and their 
similarity thus appears to be accidental. Using the scaling relations $2\beta/\nu-\eta = d_\text{eff}-2$ [assuming Eq. \eqref{eq:z_eta_Ansatz}] and $\gamma/\nu=2-\eta$, 
it follows that the similarities in 
$\beta/\nu$ and $\gamma/\nu$ for the Ising and $XY$ models are as coincidental as they are for $\eta$, both in 2D and 3D. It appears that these coincidences 
persist in all dimensions between 2 and 3. There is good reason to expect that the same also holds in the sub-Ohmic regime $s <1$. Such values 
of $s$ increase the effective dimensionality beyond 3, eventually driving all exponents to their universal mean-field values at the upper critical dimension. 
  
We next comment on other values of $q$, and how our results apply to those cases. The Ising and $XY$ models represent extreme cases of $Z_q$ clock models, 
with $q=2$, $q=\infty$, respectively. The partition 
function for the $q=4$ case is simply the square of the case $q=2$, and hence they give identical results. For larger $q > 4$, anisotropy is irrelevant,
\cite{Elitzur-Pearson-Shigemitsu_clock_model} and we thus expect the results of $U(1)$ to emerge. We therefore conjecture that the results of this paper 
for $z$, $\eta$, $\beta/\nu$, and $\gamma/\nu$, are valid for all $Z_q$ clock models. The only possible exception is the case $q=3$, also equivalent to the 
three-state Potts model, where the anisotropy with respect to a U(1)-symmetric model is known to be relevant. Although we have not checked 
this, it may still be possible that Eq. \eqref{eq:z_eta_Ansatz} holds also for a dissipative $Z_3$ clock model, 
at least for $s > 1$.\cite{footnote2}

An alternative perspective on this, supporting the notion that the scaling relation $z=(2-\eta)/s$ is valid for all $q$, may be provided by the following 
qualitative argument. The variation of $z$ with the parameter $s$ determining the range of the dissipation, expresses a variation in the effective space-time dimensionality 
of the system. This is determined by the interaction of the spins at a given site in the imaginary-time direction. Due to the long-range character of this interaction, each 
spin interacts with a large number of fluctuating copies of itself along a chain in the imaginary-time direction. Due to the summation over many spins at different Trotter slices, 
the discrete nature of the spins in a $Z_q$ clock model is washed out, even in the case $q=2$. Therefore, the manner in which the dissipation affects the effective dimensionality
does not depend on whether the spins at each space-time lattice point take on discrete or continuous values.

\section{Summary}\label{sec:summary}

We have performed Monte Carlo simulations on a generalized spin-boson model in one spatial dimension featuring non-Ohmic site dissipation and two variants of order parameter 
symmetry, namely Ising-like and $U(1)$. By tuning the imaginary-time decay exponent of the dissipative interaction, $s \in [1,2]$, we are able to continuously vary the 
effective dimensionality of the system. Apparently, the order parameter symmetry has very little bearing on the evolution of the effective dimensionality, 
$d_{\mathrm{eff}} = d + z$, as a function of the decay parameter $s$. While naive scaling estimates a crossover from criticality dominated by the dissipation 
term to an irrelevant dissipation term at $s=2$, we measure exponents in relatively good correspondence with the underlying, short-range interacting model at 
a somewhat lower value $s\approx 1.75$. Our results also suggest that for $1 \leq s \leq 2$, the exponents $z$ and $\eta$ to a good approximation obey the scaling 
relation $z = \text{max} \left\{ (2-\eta)/s, 1 \right\}$.

\acknowledgments
A.S. was supported by the Norwegian Research Council under Grant No.  205591/V30 (FRINAT). I.B.S. and 
E.B.S thank NTNU for financial support. The work was also supported through the Norwegian consortium
for high-performance computing (NOTUR). Useful communications with E. V. Herland, D. A. Huse, F. S. 
Nogueira, and Z. Te\v{s}anovi\'{c} are acknowledged.


\begin{thebibliography}{41}
\bibitem[{(Footnote)}]{footnote}
To be more specific, we carried out the joint fit for system sizes in the interval $L=57$ to $L=607$ for three data sets $\alpha =$ 0.05, 0.1, 0.2, where only 
the exponents $z$ and $\omega$ were constrained to be equal for all data sets. The value of the correction-to-scaling exponent was found to be $\omega = 0.42(25)$. The 
values of the correction amplitudes $b$ were all negative within statistical uncertainty and decreasing with increasing $\alpha$. This implies that $\alpha$ would have 
to be set to a vanishingly small value to reduce the corrections to scaling for $s=1.875$, and also justifies using relatively low values of $\alpha$ for the largest values of $s$.
\bibitem[{(Footnote)}]{footnote2}
We expect that a three-state Potts model with effective dimensionality $d_\text{eff}$ may have a first-order phase transition for $d_\text{eff} \geq 3$.
\end{thebibliography}
\end{document}